\documentclass[12pt]{article}  

\begin{document}
\begin{titlepage}
\title{ Asymptotic behaviour of a class of inhomogeneous scalar field
cosmologies }

\author{ J. Ib\'a\~{n}ez and I. Olasagasti  \\
Dpto. F\'{\i}sica Te\'orica, Universidad del Pa\'{\i}s Vasco, Bilbao, Spain. }
\maketitle
\vskip 2cm
\begin{abstract}

The asymptotic behaviour of a  class of inhomogeneous
scalar field cosmologies with a Liouville type of potential is studied. We
define a set of new variables for which the phase space of the system of
 Einstein equations is bounded. This allows us to perform a complete
analysis of the evolution of these  cosmologies. We also discuss the extension
of the cosmic no-hair theorem.    
\end{abstract}
\end{titlepage}


\section*{I\hspace{0.5cm} Introduction}
Since the proposal by Misner \cite{m} of the ``chaotic cosmology program'',
the idea that the Universe emerged from a highly irregular
state and that the inhomogeneities and the anisotropies were
washed away giving place to a highly symmetric universe, has
been one of the most attractive ideas in cosmology. In spite of
the theorem proved by Collins and Hawking \cite{ch} which
states that only a subclass of measure zero of the space of
homogeneous solutions approach isotropy and the alternative
approaches that postulated, according to the second law of
thermodynamics \cite{bp}, that the universe began in a highly
regular state, Misner's idea has come back due to the succes of the inflationary
scenarios. 

The cosmic ``no-hair'' theorems of Wald \cite{w} for homogeneous models and
of Jensen and Stein-Schabes
\cite{js} for inhomogeneous spacetimes  pointed out how the introduction of a
cosmological constant, which can be considerded as induced by a  scalar field
(inflaton), allows the  models to 
isotropize approaching  the de Sitter solution. This situation, however, may
dramatically change  if one takes into account the dynamical behaviour of the 
scalar
field. Heusler \cite{h}, for example, extending the theorem of
Collins and Hawking to the case of convex and positive
potentials of the scalar field possessing a   local minimum has
shown that only the homogeneous Bianchi  models which admit a
FRW solution as a particular case approach isotropy. Later Kitada
and Maeda \cite{km}  and Ib\'a\~nez et al. \cite{i}  have shown that when one
assumes  a Liouville type of potential for the scalar field (exponential
potential), the Wald's theorem for homogeneous solutions still applies if the
exponential potential is quite flat. 

Although Jensen and Stein-Schabes \cite{js} extended the Wald theorem to 
inhomogeneous
solutions, little is known about the effect of the dynamics  of
the scalar field on the asymptotic behaviour of the models. The first
attempts to incorporate the effect of the dynamical evolution of the scalar 
field was
made by Goldwirth and   Piran \cite{gp} using numerical studies of inhomogeneous
models and, later on by Calzetta and Sakellariadou \cite{cs} by studying the 
evolution
of a family of inhomogeneous Cauchy data. The first inhomogeneous
scalar field exact solutions with exponential potential of the Einstein
field equations were obtained by Feinstein and Ib\'a\~nez \cite{fi}  and
it was shown there that  the scalar field does not guarantee by itself
that the model inflates or isotropize. Although there are some exact
inhomogeneous solutions obtained up to now, we are lacking a result similar
to Heusler's. In particular, for the exponential potential, it would be 
interesting
to study the conditions on the scalar field leading to inflation and 
isotropization.

The asymptotic behaviour of homogeneous but anisotropic solutions for a 
perfect fluid
has been widely studied. By  using the kinematical quantities of the fluid as
variables, the field equations can be written as an autonomous system of 
ordinary
differential equations 
\cite{hw}. One of the most distinctive features is that the equilibrium 
points of
these systems are self-similar solutions. This analysis has recently been 
extended to
the case when the matter source is a scalar field with exponential potential 
\cite{vh}.
Since the scalar field is homogeneous, one can globally associate  with it a 
perfect
fluid, and then by using the kinematical variables of the fluid,  the Einstein
field equations  decouple and the phase space of the system becomes bounded. 

The main difficulty in analyzing the
asymptotic behaviour  of inhomogeneous solutions is related with the fact that
the field equations involve  partial derivatives. The study of such
systems was initiated in 
\cite{hwg}, where a particular family of self-silmilar solutions with
perfect fluid, for which the Einstein equations  reduced to an autonomous
system, was studied. Since the source of the metric was a perfect
fluid, the kinematical quantities could be used again  to analyse the asymptotic
behaviour of the system, as in the homogeneous case. 

In dealing with an 
inhomogeneous scalar field one can not apply the analysis developped in
\cite{hwg}  due to the fact that one can not globally interpret the
scalar field as a perfect fluid. Therefore in order to investigate  whether the
inhomogeneous scalar field cosmologies undergo an inflationary epoch leading 
to the
homogenization and the isotropization of the spacetime, one does have to look 
for a
different way to tackle this problem. 

In  this paper we initiate the study of the
asymptotic behaviour of the scalar field cosmologies by considering, as a 
first step, a 2 parameter family of
$G_2$ self-similar solutions. The fact
that the scalar field is not equivalent to a perfect fluid prevents the use
of the kinematical quantities to describe the evolution of the
solutions. Despite of the lack of a preferred timelike congruence in the
spacetime we manage to find a set of new variables, in terms of which the phase
space of the system becomes  bounded. This allows us to perform a complete 
analysis
of the asymptotic behaviour of these spacetimes. In addition, the way to
introduce the set of new variables is a promising method for dealing
with  more general solutions.     

The plan of the paper is as follows: in Section 2 we present the metric
and  introduce the new variables. Section 3 is devoted to the analysis
of the phase space and in Section 4 we discuss the results.           

\section*{II\hspace{0.5cm} The metric and the compactified phase space}

We will consider solutions with one-dimensional inhomogeneity. These metrics are
described by the generalized Einstein-Rosen spacetimes which admit an Abelian 
group
of isometries $G_2$. If the two Killing vectors are hypersurface orthogonal the
line element can be written as
\begin{equation}
ds^2=e^F\,\left( -dt^2+dz^2\right) +G\,\left( e^h\,dx^2+e^{-h}dy^2\right),
\end{equation}
where the metric functions depend on $t$ and $z$ and the Killing vectors are
$\partial_x$ and $\partial_y$.

The matter source for the metric is that of a minimally coupled scalar field 
with potential, for which the stress-energy tensor is given by
\begin{equation}
T_{ab}=\bar\phi_{,a}\bar\phi_{,b}-g_{ab}\left( \frac{1}{2}
\bar\phi_{,c}\bar\phi^{,c}+V(\bar\phi)\right),
\end{equation}
(latin indices run from 0 to 3) with the Liouville type of the potential 
\begin{equation}
V(\bar\phi)=\Lambda\,e^{k\bar\phi}, \quad \Lambda\geq 0.
\end{equation}
It is well known \cite{tt} that as long as the gradient of the scalar field 
remains
timelike, (2) can be rewritten as a perfect fluid stress-energy tensor. 

To simplify the equations we will concentrate in this paper on the class of 
solutions
for which the element of transitivity surface $G$ is homogeneous
\begin{equation}
G=G(t),
\end{equation}
which is suitable to a description of cosmological models.

It was shown in \cite{fil} that if one assumes separability of  the metric  and
separability, in the additive sense, of the scalar field  one obtains, from 
the field
equations, that the dependence of these functions on the variable $z$ is linear
\begin{equation}  
h(t,z)=p(t)+az,\quad F(t,z)=f(t)+cz,\quad \bar\phi(t,z)=\phi(t)+bz,
\end{equation}
where $a$, $b$ and $c$ are arbitrary constants which drive the inhomogeneity.
This linear behaviour for the inhomogeneity has been recently considered, in
a different context, by Vilenkin \cite{v}.  The general solution for this 
class of
metrics and the study of few particular examples were given in \cite{fi}. 

Before going to the field
equations it is worth mentioning  that the metric (1)-(5) admits a
homothetic vector field given by
\begin{equation}
{\bf H}=\frac{2}{c}\,\frac{\partial}{\partial{\bf z}}+\left(
1-\frac{a}{c}\right)
\,x\,\frac{\partial}{\partial{\bf x}}+\left( 1+\frac{a}{c}\right)
\,y\,\frac{\partial}{\partial{\bf y}}. 
\end{equation}

It was conjectured in \cite{hwg} that the self-similar solutions could be the
attractors of the $G_2$ cosmologies and in this sense the metric (1)-(5) would
play an
important role in the study of the asymptotic behaviour of the inhomogeneous
solutions. The metric (1)-(5) is similar to  the  metric 
studied by Hewitt {\it et al} \cite{hwg} (it is obtained interchanging $t$ by
$z$ in (5)). Both metrics, however, differ   in the
character of the orbits of the similarity group and in the type of the source.
In our case the orbits of (6) are spacelike given by
$t=$constant and the scalar field is inhomogeneous and, therefore, it is not
equivalent, in general, to a perfect fluid. 

The Einstein equations and the Klein-Gordon equation for the scalar field
with the ansatz (5) are given by
\begin{eqnarray}
\frac{\ddot G}{G}-2\,V(\phi)\,e^F & = & 0\: , \\
\ddot p+\frac{\dot G}{G}\,\dot p & = & 0\: , \\ 
a\dot p-c\frac{\dot G}{G}+2b\dot\phi & = & 0\: , \\ 
\frac{\ddot
G}{G}-\frac{1}{2}\,\left(\frac{\dot G}{G}\right)^2-\dot f\frac{\dot
G}{G}+\frac{1}{2}\dot p^2+\dot\phi^2+\frac{1}{2}a^2+b^2 & = & 0\: , \\ 
\ddot f
-\frac{1}{2}\,\left(\frac{\dot G}{G}\right)^2+\frac{1}{2}\dot
p^2+\dot\phi^2-\frac{1}{2}a^2-b^2 & = & 0\: , \\ 
\ddot\phi+\frac{\dot
G}{G}\dot\phi+kV(\phi)e^F & = & 0\: , 
\end{eqnarray}

For a massless scalar field, $V=0$, the former system of equations turns out
to be easily integrable giving the following solution
\begin{eqnarray}
p(t)&=&A\ln t, \qquad \phi (t)=B\ln t, \qquad G(t)=t,  \nonumber \\
f(t)&=& \left(\frac{1}{2}A^2+B^2-\frac{1}{2}\right)\ln t+\frac{1}{2}\left(
\frac{1}{2}a^2+b^2\right)\,t^2,
\end{eqnarray}  
where $A$ and $B$ are constants subject to the condition
\begin{equation}
aA-c+2bB=0\: .
\end{equation}

When $V\neq 0$, from (7) or (12), and due to the exponential form of the
potential 
\begin{equation}
c=-kb\: .
\end{equation}
In this case if the constant $b=0$ then, from (9), either
$\dot p=0$ and the metric is Bianchi type VI$_0$, or $a=0$ and the metric is
Bianchi type I. Let's note that whatever value $V$ takes, when $a=b=0$ the
metric is Bianchi type I.    

In order to look for a suitable set of new variables to compactify the phase 
space
 we write the generalized Friedman equation which is obtained from
(7) and (10)
\begin{equation}
-\frac{1}{2} \left( \frac{\dot G}{G}\right)^2-\dot f\frac{\dot G}{G}+
\frac{1}{2}\dot
p^2+\dot\phi^2 +\frac{1}{2}a^2+b^2+2V(\phi)e^F=0.
\end{equation}
This equation suggests the introduction of the following set of variables
\begin{eqnarray}
\beta=\frac{\dot p}{\dot G/G+\dot f}\, , & &\quad\Psi=\sqrt{2}\frac{\dot\phi}
{\dot G/G+\dot f}\: , \nonumber \\
\Phi=\frac{\dot f}{\dot G/G+\dot f}\, , & &\quad\Gamma=\frac{2\sqrt{Ve^F}}{\dot
G/G+\dot f}\: .
\end{eqnarray}
In terms of these new variables (16) is written as
\begin{equation}
1-\beta^2-\Psi^2-\Phi^2-\Gamma^2=\frac{\left( a^2+2b^2\right)}{\left(\dot G/G+
\dot f\right)^2} \geq 0\: .
\end{equation}
Therefore the space of solutions described by the variables (17) is
bounded to the inside of the 4-sphere (18). Let's note that, from  (16), $\dot
G/G+\dot f$ has to be different from zero, unless $\Lambda$ be negative.   

In dealing with perfect fluid models new variables were defined as the
kinematical quantities of the fluid divided by an appropiate power of the rate
of expansion $\theta$. This assures a good behaviour of these variables near
the initial singularity. Let's note that our new variables (17) are
divided by the quantity $\dot G/G+\dot f$ which is not related with the
expansion of any timelike congruence. Nevertheless, from the general behaviour
near
the initial singularity found by Belinskii {\it et al} \cite{bkl} and from the
work of Isenberg and Moncrief \cite{im}  we can
assume that the metric (1)-(5), near the initial singularity behaves, for each
value of the coordinate $z$, like a Kasner model. Therefore, 
when $t\rightarrow 0$
\begin{equation}
G\sim t,\qquad f\sim p\sim \phi\sim \ln t\: ,
\end{equation}
and the variables (17) remain bounded when $t$ tends to zero. 

 From (18) we  see that    
the points on the surface of the
4-sphere represent either homogeneous Bianchi type I solutions (when constants
$a$ and $b$ are zero) or the initial singularity  of the models. 

By using (17),  (9) is written as
\begin{equation}
a\beta+b \sqrt{2}\Psi-c(1-\Phi)=0\: .
\end{equation}
Except for the trivial case when $a=b=0$, this equation gives a constraint
for the constants $a$ and $b$. Alternatively, if $a$ and $b$ are fixed one
can look at this equation as giving a plane,   intersection of which with the
sphere (18) describes the phase space. In the study of the equilibrium
points in the next Section  $a$ and $b$ will be  arbitrary chosen constrained by
(20).

Using the variables (17), (7) becomes
\begin{equation}
\frac{\ddot
G}{G}=\frac{1}{2}\,\left(\frac{\Gamma}{1-\Phi}\right)^2\,\left(\frac{\dot
G}{G}\right)^2.
\end{equation}
This equation decouples from the rest of field equations if we  introduce a new
time coordinate
\begin{equation}
\frac{d\tau}{dt}=\frac{\dot G}{G}+\dot f=\frac{1}{1-\Phi}\,\frac{\dot G}{G}\: .
\end {equation}
Near the initial singularity 
\begin{equation}
\frac{d\tau}{dt}\sim\frac{1}{t}\quad \Rightarrow\quad \tau\rightarrow 
-\infty\: ,
\end{equation}
thus $\tau$ varies from $-\infty$ to $+\infty$. In terms of this new time  the
field equations  are written as
\begin{eqnarray}
\beta' & = & -\beta\left( 1-\beta^2-\Psi^2-\Phi^2\right)\: , \\
\Psi' & = & -\Psi\left(
1-\beta^2-\Psi^2-\Phi^2\right)-\frac{k}{2\sqrt{2}}\,\Gamma^2\: ,
\\
\Phi' & = & (1-\Phi)\,\left( 1-\beta^2-\Psi^2-\Phi^2\right)-\frac{1}{2}\,
\Gamma^2\: ,
\\
\Gamma' & = & -\Gamma\left(
-\beta^2-\Psi^2-\Phi^2+\frac{1}{2}\Phi-\frac{k}{2\sqrt{2}}\,\Psi\right)\: ,
\end{eqnarray}
where $'$ means derivative with respect the new time $\tau$.
Differentiating (20) with respect to $\tau$ one can easily see that  the
constraint equation (20) holds for all values of $\tau$, as long as the
initial conditions verify the equation (20). Threrefore (24)-(27) along
with the constraint equation (20) describe the evolution of the metric
(1)-(5).         

\section*{III\hspace{0.5cm} The Equilibrium Points and The Invariant Sets}

In this Section we shall study the qualitative behaviour of the trajectories 
of the
system (24)-(27). Let's first note that the system admits the discrete symmetry
$\Gamma\rightarrow -\Gamma$, and, therefore, without loss of generality the
study of the equilibrium points will be restricted to $\Gamma\geq
0$.

\subsection*{Equilibrium points}

The equilibrium points of an autonomous system play an important role in 
describing the qualitative behaviour of its solutions. The local 
stability of the equilibrium
points are given by the eigenvalues of the linearized differential equations. 
The equilibrium points of the system (24)-(27) can be found explicitly 
and we now give them and their character.

\vskip 1cm
\noindent
I $\displaystyle \qquad \left\{\beta=0\quad \Psi=-\frac{k\sqrt{2}}{2+k^2}\quad
\Phi=\frac{k^2}{2+k^2}\quad \Gamma=\frac{2\sqrt{2}}{2+k^2}\right\}$ \\

 This point is outside  the sphere when $k^2<2$ and   it is on the 
surface of the sphere when $k^2=2$. The constraint equation (20) is 
trivially satisfied for all values of the constants $a$ and $b$. The
corresponding metric and the scalar field when $k^2>2$  are
\begin{eqnarray}
ds^2 & = & Ce^{k^2At/2-kbz}\left(-dt^2+dz^2\right)+e^{At}\left(e^{az}dx^2+
e^{-az}dy^2\right)\: , \nonumber \\
\phi & = & -\frac{k}{2} At+bz\: ,
\end{eqnarray}
where
\begin{equation}
C=\frac{A^2}{\Lambda}\: , \qquad
A=\sqrt{\frac{2\left(a^2+2b^2\right)}{k^2-2}}\: . 
\end{equation}

This metric was  obtained in \cite{fi} and is of Bianchi type VI.
The eigenvalues of the linearized system are
\begin{equation}
\frac{-2}{k^2+2},\quad \frac{-2}{k^2+2},\quad
\frac{-1+\sqrt{5-2k^2}}{k^2+2}, \quad \frac{-1-\sqrt{5-2k^2}}{k^2+2}.
\end{equation}
When $k^2<5/2$ the equilibrium point is a stable node and when $k^2>5/2$  a
stable focus.

\vskip 1cm
\noindent
II $\displaystyle \qquad
\left\{\beta=0\quad\Psi=-\frac{k}{2\sqrt{2}}\quad\Phi=\frac{1}{2}\quad\Gamma=
\frac{\sqrt{6-k^2}}{2\sqrt{2}}\right\}$ \\

This equilibrium point is on the surface of the sphere. It disappears when
$k^2>6$ and  coincides with the former equilibrium point when $k^2=2$.
Substituting the former values of the variables of the equilibrium 
point  into (17) one easily gets that the solution  corresponds to a 
homogeneous Bianchi type I solution
($a=b=0$). The line element and the scalar field are 
\begin{eqnarray} 
ds^2 & = & C
t^m\left(-dt^2+dz^2\right)+t^m\left(dy^2+dz^2\right),
\qquad m=\frac{4}{k^2-2}\: ,
\nonumber \\ 
\phi & =& -\frac{2k}{k^2-2}\,\ln t \, ,
\end{eqnarray}
where
\begin{equation}
C=\frac{2\left(6-k^2\right)}{\Lambda\left(k^2-2\right)^2}\: ,
\end{equation}
and the metric represents the FRW universe with massless minimally coupled 
scalar field as a source. The eigenvalues of the linearized system are
\begin{equation}
-\frac{1}{8}(6-k^2),\quad -\frac{1}{4}(2-k^2),\quad -\frac{1}{8}(6-k^2),\quad
-\frac{1}{8}(6-k^2)\: .
\end{equation}
When $k^2<2$ this equilibrium point is a stable node  but when $2<k^2<6$  is a
saddle point.

It is important to note that the constraint equation (20) is satisfied 
not only because $a$ and $b$ vanish but because $\beta=0$ and $\sqrt{2}\Psi+
k(1-\Phi)=0$. If we look into the evolution in time of a particular solution
with  $a$ and $b$ different from zero, and with $k^2<2$, $\beta$, $\Psi$ and
$\Phi$ will take values such that the constraint equation will be always
satisfied and as $t\rightarrow\infty$ the solution will approache this
equilibrium point becoming, therefore, homogeneous and isotropic.    

\vskip 1cm

\noindent
III $\displaystyle \qquad \left\{\beta^2+\Psi^2+\Phi^2=1,\quad 
\Gamma=0\right\}$ \\
   
This ring of equilibrium points belongs to  the
surface of the sphere and  represents, therefore, Bianchi type I 
solutions with a minimally coupled massless scalar field. By choosing a 
particular point on the surface,
i.e. $\beta=a_1, \Psi=a_2$ and $\Phi=a_3$ with $a_1^2+a_2^2+a_3^2=1$,
the eigenvalues of the linearized system are
\begin{equation}
\frac{1}{2}\left(2-a_3+\frac{k}{\sqrt{2}}a_2\right),\quad 2(1-a_3),\quad
0,\quad 0.
\end{equation}
Hence, any of these points are unstable, except when
$\Phi=1$ $(a_3=1)$, which corresponds to the Minkowski spacetime. 

\subsection*{Invariant sets}

Besides the equilibrium points, the existence of invariant sets help to 
describe the qualitative behaviour of the solutions of an autonomous 
system. In our case, there are three invariant sets, two of them 
describing massless and exponential potential
scalar field solutions respectively while the third gives the Bianchi type I
solutions. 

The points with $\Gamma=0$ (massless scalar field spacetimes) compose an
invariant set of the dynamical system (24)-(27) whose solutions are given by
(13) and (14). The dynamics of the points of this subspace can be easily
studied: when $\Gamma=0$ the dynamical system reduces to
\begin{eqnarray}
\beta' & = & -\beta\left(1-\beta^2-\Psi^2-\Phi^2\right)\: , \nonumber \\
\Psi' & = & -\Psi\left(1-\beta^2-\Psi^2-\Phi^2\right)\: , \\
\Phi' & = & \left(1-\Phi\right)\left(1-\beta^2-\Psi^2-\Phi^2\right)\: ,\nonumber
\end{eqnarray}
with  straight lines as solutions 
\begin{equation}
\Phi=\frac{\Phi_0-1}{\beta_0} \beta+1,
\qquad \Psi=\frac{\Psi_0}{\beta_0}\beta\: .
\end{equation}
These lines start on the surface of the sphere $\beta^2+\Psi^2+\Phi^2=1$ 
and intersect at the point $\beta=\Psi=0, \Phi=1$. Thus the solutions 
evolve from the Kasner initial singularity to the Minkowski spacetime.

The second invariant set is given by the points with $\Gamma>0$ (and by symmetry
$\Gamma<0$), with equilibrium points being the points I and II described above.

Finnally the third invariant set is given by the points on the  
surface of the sphere ($\beta^2+\Psi^2+\Phi^2+\Gamma^2=1$). In this  case the
dynamical system is reduced to
\begin{eqnarray}
\beta' & = & -\beta\left( 1-\beta^2-\Psi^2-\Phi^2\right)\: , \nonumber \\
\Psi' & = & -\left(\Psi+\frac{k}{2\sqrt{2}}\right) \left(
1-\beta^2-\Psi^2-\Phi^2\right)\: , \\ 
\Phi' & = & \left( \frac{1}{2}-\Phi\right) \left(
1-\beta^2-\Psi^2-\Phi^2\right)\: . \nonumber 
\end{eqnarray}
The solutions of this system are again straight lines. 
\begin{equation}
\Phi=\left(\Phi_0-\frac{1}{2}\right)\frac{\beta}{\beta_0} +1,\qquad
\Psi=\left(\Psi_0+\frac{k}{2\sqrt{2}}\right)\frac{\beta}{\beta_0}-
\frac{k}{2\sqrt{2}}\:
, 
\end{equation}
and the
equilibrium points of the system (37) are the points II and III. 

The behaviour of the system (24)-(27) can be visualized in  Fig.1 where 
the phase
space for $\beta=0$ and $\Gamma\geq 0$ is depicted for different values of the
constant $k$. The positions of the equilibrium point I as function of $k$ are
represented by the dashed line. All the solutions start on the circle
$\Phi^2+\Psi^2=1, \Gamma=0$. When $\Gamma=0$ (massless scalar field), 
for all the cases, the solutions tend to the point $\Phi=1, \Psi=0$.  
When $k^2<2$ the point I is
outside the sphere  and the solutions tend to the equilibrium point II. 
When $2<k^2<6$ point I is inside the sphere and the solutions evolve 
either to the point I (those that go
inside the sphere) or to the point II (those that lie on the boundary of the
sphere). When $k^2>6$ the only attracting point is point I (inside the 
sphere). The
trajectories on the surface start and finish on the circle $\Phi^2+\Psi^2=1,
\Gamma=0$ in such a way that their projections on the plane $\Gamma=0$ are
straight line directed to the point
$\Psi=-\frac{k}{2\sqrt{2}}\Phi=\frac{1}{2} $ which is outside the surface.

 The
behaviour described above remains the same when $\beta$ is different 
from zero.

\section*{IV\hspace{0.5cm} Conclusions}
We have studied in this paper the asymptotic behaviour of a
particular class of inhomogeneous solutions with a minimally
coupled scalar field with an exponential potential. The metric
belongs to the class of $G_2$ cosmologies. 

We have succeeded to define a set of new variables for which the entire
phase space is bounded by  a 4-sphere. From the analysis of the dynamical system
obtained for this metric, we deduce the following:

i) As in the homogeneous case \cite{vh}, the dynamical behaviour of the
metric depends on the parameter $k$ which is related to the mass of
the scalar field. The parameters $a$ and $b$ which drive the
inhomogeneity do not play a significant role in this
behaviour.

ii) When $k^2<2$ the only equilibrium point is that given by the FRW
universe. The trajectories evolve from the surface of the 4-sphere, 
representing the Kasner regime near the initial singularity, towards the 
isotropic equilibrium point II. When $k^2>2$ there are two equilibrium 
points, one is the homogeneous Bianchi type VI
(point I) and the second is the FRW solution which is a saddle point 
and, therefore, unstable against small changes of the initial conditions. 
This means that the cosmic no-hair theorem holds in this case provided 
$k^2<2$. 

The behaviour described above is almost identical to that of homogeneous models
\cite{vh}, indicating that the introduction of the inhomogeneity through 
the constants
$a$ and $b$ does not affect remarkably the dynamical system. It is
likely that the presence of a more ``strong'' inhomogeneity, assuming
another dependence on the spatial coordinate, may change the behaviour
of the solutions and the conditions to isotropize. We consider,
therefore, that it is important to extend this analysis to a more
general  class of inhomogeneous scalar field solutions.  
 
As to the inflation of the models, for inhomogeneous solutions there is no 
natural way to see whether the models inflate. The hypersurfaces
$\bar\phi=$constant do not define a globally timelike observer and, therefore,
one should use a weaker way to specify the inflationary behaviour. It has been
suggested \cite{fir}, for example, to look at the fulfillment of the strong
energy condition, the breaking of which is a necessary condition for a model to
inflate. The energy density and the pressure of the scalar field are given by
\begin{equation}
\rho=-\frac{1}{2}\bar\phi_{,a}\bar\phi^{,a}+V, \quad
p=-\frac{1}{2}\bar\phi_{,a}\bar\phi^{,a}-V.
\end{equation}
In terms of the new variables, the breaking of the energy condition is 
written as \begin{equation}
3p+\rho=e^{-F}\left(\frac{\dot G}{G}+\dot
f\right)^2\left[\Psi^2-\frac{1}{2}\Gamma^2-\frac{2b^2}{\left(\dot G/G+\dot
f\right)^2}\right]<0.
\end{equation}
For the equilibrium points I and II the expression 
$\Psi^2-\frac{1}{2}\Gamma^2$ turns out to be
\begin{equation}
\frac{2(k^2-2)}{(2+k^2)^2} \qquad {\rm and}\qquad \frac{3(k^2-2)}{16} 
\end{equation}
respectively. That means that if $k^2<2$ both points describe spacetimes 
inflating,
but if $k^2>2$ the energy condition could be broken  only by the solution
represented by the point I depending on the value of $b$. This behaviour is
similar again with that of the homogeneous models \cite{i}.    

\vskip 1cm
\noindent
{\bf Acknowledgments}

\noindent
J.I. is grateful to Prof. A.A.Coley and Dr. R. van den Hoogen for stimulating
discussions. We also want to acknowledge to Prof. A.Feinstein for his 
comments.This work is
supported by a grant DGICYT PB93-0507. I.O. 's work is supported by
a fellowship from the DGICYT ref. FP94.

\pagebreak

\noindent
{\bf Figure Captions}
\vskip 1cm
\noindent
Figure 1.- Phase space when $\beta=0$ and $\Gamma\ge 0$, and for 
different values of
the constant $k$. The equilibrium point I is represented by a square and the
equilibrium point II by a circle. The dashed line describes the position 
of the point I for different values of
$k$. Faded lines represent the trajectories of the solutions lying on 
the surface of
the sphere which is an invariant set. The plane $\Gamma=0$ is an invariant set
describing the massless scalar field solutions.

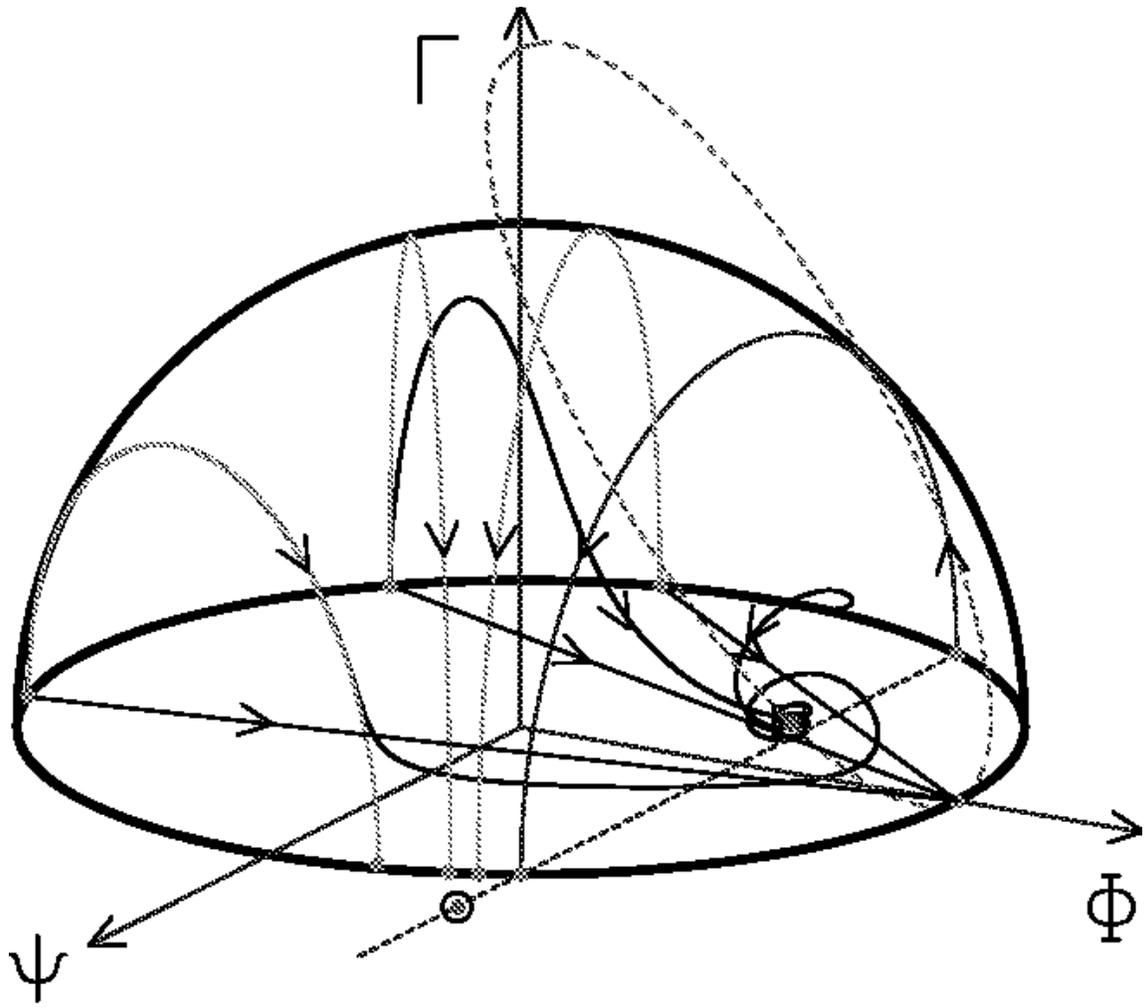
\end{document}